\documentclass [12pt]{article}
\usepackage {graphicx}

\begin{document}

\title {Distant source image motion due to gravitational field of
the Galaxy stars}

\author{V.I.Zhdanov, E.V.Fedorova, A.N.Alexandrov \\
\begin{small} \it National Taras Shevchenko University of Kyiv
\end{small}
}

\date{}

\maketitle

\begin{abstract}
Gravitational field of stars of the Galaxy causes additional
motions of images of extragalactic sources. For typical
realization of stars this results in a small  rotation of
extragalactic reference frame in the direction of Galactic
rotation. We estimate angular velocity of corresponding image
motion and show that it is different in case of discrete and
continuous matter distribution in the Galaxy.
\end{abstract}

\section{Introduction}

Perspectives of microarcsecond astrometry raise questions
concerning motion of images of extragalactic radiation sources due
to gravitational field of moving stars
\cite{Kardashev},\cite{ZZ},\cite{Zh},\cite{Sazhin},\cite{SZVK}.
This gravitational image motion (GIM) may be comparable to proper
motion of quasars. There is some hope that corresponding accuracy
to observe such effects will be available for future space-based
radio interferometers \cite{Kardashev}. However, it will be
difficult to separate GIM from real proper motions, and this puts
some limit for accuracy of fundamental reference frame based on
extragalactic sources \cite{Sazhin},\cite{SZVK}. Indeed, we cannot
take into account positions of all stars in the Galaxy; we only
may work with probability distribution of Galactic stars leading
to distribution of image motions of certain extragalactic object
\cite{ZZ},\cite{Zh}.

It has been pointed out in \cite{ZZ},\cite{Zh} that stochastic GIM
of an extragalactic source induced by stellar motions is
accompanied with a systematic component which depends upon bulk
velocity of microlensing stars, including stars that are far away
from the line of sight. This requires some explanation. If we
perform a statistical averaging using all possible positions of
microlenses (stars), we shall obtain a value different from the
most probable value of GIM that might be observed at the present
epoch. There is also a difference between the statistical average
and arithmetic mean over all reference sources used in
extragalactic reference frame. This difference appears to be
essential (an example will be given in this paper below). The
reason is that in reality the number of these sources is not
sufficiently large. In this paper we treat this question in more
detail. We show that one can estimate probable GIM effect, without
recourse to probability distributions, after some restriction of
sample of observed sources. The result differs in case of
continuous (e.g., dark matter) and discrete (e.g., stars) mass
distributions.

\section{Motion of source image microlensed by a point mass}

In case of microlensing of quasars by Galaxy stars we may neglect
cosmological curvature. Evidently this does not contradict to assumption
that the radiation source is at the infinity.

Let unperturbed light ray moves from the infinitely distant radiation source
in negative direction of $z$-axis of Cartesian coordinates
{\{}$x,y,z${\}}, the observer being at the origin. Let position of
microlensing point mass $M$ be $({\bf r},z)$, where ${\bf
r}=(x,y)$ is a two-dimensional vector in the transverse plane;
i.e. $r=\vert {\bf r}\vert$ is the impact distance of the
unperturbed ray with respect to the mass. We consider the case of
weak microlensing $r>>(mz)^{1 / 2}$, $m=GM/c^2$; therefore the
light trajectory in the gravitational field of the point mass can
be obtained in the post-Newtonian approximation. Note that in our
problem we may have $r$ of the same order as $z$. In this case

\begin{equation}
\label{eq1} {\bf \Psi}  = - 2 \frac{\bf r}{r^2}\left[1 + z (z^{2} +
r^{2})^{-1 / 2}\right]  .
\end{equation}

\noindent where ${\bf \Psi}  = (\Psi _1 ,\Psi _2 )$ is a two
dimensional vector describing the source image angular shift per
unit value of $m$ \cite{Zh},\cite{Sazhin},\cite{Pyr}. Formula
(\ref{eq1}) may be obtained from a more general relation
\cite{Pyr} in the case of an infinite source.

Further $ {\bf V}_p =({\bf v},w)$ stands for velocity of the point
microlens, $w$ is the velocity component parallel to the line of
sight and ${\bf v}$ represents the transversal components. In
virtue of (\ref{eq1}) this leads to the source GIM that equals to
$m {\bf U}$ (in radians per unit of time), where \cite{Zh}

\[
{ {\bf U}} = \frac{d {\bf \Psi } }{dt} = - 2\left\{
{\frac{1}{r^4}} \left[ {\bf v}r^2 - 2 {\bf r}( {\bf r} \cdot
{\bf v})\right] \cdot \left[1 + \frac{z}{\sqrt {z^2 + r^2} }\quad +\right.\right.
\]

\begin{equation}
\label{eq2}
+\left.\left.\frac{zr^2}{2(z^2 + r^2)^{3 / 2}}\right] +  {\frac{w{\rm {\bf r}}
- z{\rm {\bf v}} / 2}{(z^2 + r^2)^{3 / 2}}} \right\}
\end{equation}

Here ${\bf U}$ is a function of the microlens position $({\bf
r},z)$ and its velocity ${\bf v}$.

\section{Motion of source image microlensed by Galaxy stars}

Because of smallness of the effect, the action of all Galaxy stars will be
taken into account linearly. This will be performed by integrating (\ref{eq2}).
However this requires justification. In fact we are dealing with extended
sources; this is not taken into account by Eq.(\ref{eq2}). As we shall see below,
this may be important in our problem.

We are interested in some estimate of probable GIM value that may be
obtained after taking arithmetic mean for as many reference sources as
possible. As we pointed out in the Introduction, this is not a statistical
average, because in reality the number of observed sources is limited. For
example, the total number of extragalactic sources in ICRS is about 600.
Moreover, this is much lesser, if we confine ourselves, e.g., to the Galaxy
plane.

We deal with an ensemble of possible positions of stars in the
Galaxy. Consider a remote extended source, which has impact
distance $r_{i}$ with respect to $i$-th star in this realization.
Let $p=\min\{r_{i}\}$, where this minimum is taken over all
Galactic stars. Now we consider two type of events: $(A)$ when
$p>>L$, where $L=D \cdot \alpha _S$, $\alpha _S$ is angular size
of the source, $D$ is a typical distance to microlens (of the
order of 50 \textit{kpc}; $(B)$ when $p\sim L$ or $ p<L$ (at least
one of stars is projected onto the source). We suppose that for
typical value of $L$ we have $L<R_E=(4mD)^{1/2}$, in this case we
may strengthen our arguments by considering even larger domain
$B'$, defined by condition $p<L$ for corresponding realizations of
stars. The rest of events we refer to the domain $A'$;
corresponding probability is $P_{A' } \approx 1$. The events from
$B'$ can be separated in observations: these are strong
microlensing events that are characterized by considerable
brightness amplification and relatively fast image motions. These
events might be taken into account, including the case of an
extended source \cite{ZhSal},\cite{ZhAlSal}. Probability of these
events $P_{B' }\sim $10$^{ - 6}$ is so small that they are not
practically essential. In case of limited number of reference
sources it is reasonable to exclude such rare microlens
realizations that would hardly occur during this century.

Nevertheless, though $P_{B' }<< P_{A'}$, the input of $B'$ into
the statistical average GIM $<{\bf U}>$ is essential because the
events $B'$ introduce large image velocities. To present an
example, we derive below a nonzero value of GIM for a fictitious
observer at the center of the Galaxy. Because of stationary
rotation of the Galaxy, one must have $<{\bf U}>=0$ (no matter,
for an infinite time average or for a statistical average). This
shows that in this case the sets $A'$ and $B'$ compensate each
other. However, for a limited time interval we
typically have a nonzero GIM value, because we do not meet
realizations from $B'$.

Therefore, the question is how to obtain a consistent GIM
estimate. The most correct way is to obtain probability for all
necessary velocity intervals. However, to have an
order-of-magnitude estimate, we can avoid calculations of
probability distributions, if we confine ourselves to obtain
average value of GIM in the domain $A'$. Further we have in mind
just this average when speaking about average GIM effect. It is
important to note that this value practically does not depend upon
the exact size of $B'$. This is provided by convergence of
integrals that will be considered below.

Thus we return to Eq.(\ref{eq2}), which describes contribution of
a single star. We assume that a star at the point $({\bf r},z)$
has velocity ${\bf V}_p ({ \bf r},z)$. Then we consider sum over
all stars in the Galaxy, which must further be averaged with the
Galactic mass density $\rho ( {\bf r},z)$. This enables us to pass
on integration yielding average GIM in the domain $A'$ for the
weak microlensing events.

\begin{equation}
\label{eq3}
 < { {\bf U}}_{tot} > _{A'} = \frac{G}{c^2}\int {dz\int {d^2{ {\bf
r}}\mbox{ }\rho ({ {\bf r}},z)\mbox{ }{ {\bf U}}({ {\bf r}},z,{
{\bf V}}_p ({ {\bf r}},z))} }
\end{equation}

\noindent
where we suppose that $\rho $ vanishes outside a bounded domain.

Taking into account the explicit form (\ref{eq2}) it is easy to
see that considerable contribution may be due to stars at large
impact distances from the line of sight. Also one can show that the
singularities in (\ref{eq3}) for small $r$ are integrable. This
allows us to avoid the question about exact value of lower limit
of $r$ in the definition of the domain $A'$.

\section{Continuous versus discrete mass distribution}

It is interesting to compare GIM (\ref{eq3}) with a corresponding
expression in case of a continuous mass distribution $\rho =\rho
({\bf r},z,t)$, satisfying the continuity equation

\begin{equation}
\label{eq4} \frac{\partial \rho }{\partial t} + div(\rho {{\bf
V}}_p ) = 0.
\end{equation}

In this case Eq. (\ref{eq1}) should be integrated with the density $\rho $:

\[
 { {\bf \Psi }}_{tot} = \frac{G}{c^2}\int {dz} \int {d^2{{\bf r}}}
\rho ({ {\bf r}},z,t) { {\bf \Psi }}({ {\bf r}},z).
\]

Note that we have the same formula for ${\bf \Psi }_{tot}$ in case
of continuous and discrete matter distribution. However,this is
not true for GIM.

The shift of remote source image in case of continuous matter
distribution has been treated in \cite{MinShal}. These shifts
appear to be of the order of $10^{ - 5}\div 10^{-6}$
radians. However this value cannot be observed from the Solar
system. In order to deal with observable (in principle) values one
must derive changes of the source positions with time:

\begin{equation}
\label{eq4a}
 {\bf U}_{tot}^\ast \equiv \frac{d {\bf \Psi }_{tot} }{dt}
=\frac{G}{c^2}\int {dz} \int {d^2{\rm {\bf r}}} \mbox{
}\frac{\partial \rho }{\partial t}\mbox{ }{\rm {\bf \Psi }}({\rm
{\bf r}},z).
\end{equation}

This will be compared to  $ < {\rm {\bf U}}_{tot} > _{A'} $ .
After standard elimination of singular points at $r=0$ in ${\bf
U}_{tot}^\ast $ and taking into account Eq. (\ref{eq4}) we
represent this integral as sum of two terms

\begin{equation}
\label{eq5} { {\bf U}}_{tot}^\ast = { {\bf U}}^{(0)} + { {\bf
U}}^{(\ref{eq1})},
\end{equation}

\noindent ${\bf U}^{(0)}$and ${\bf U}^{(1)}$ being two-dimensional vectors,
the components of the first term ($i$=1,2)

\[
U_i^{(0)} = \frac{G}{c^2}\int {dz} \int {d^2{ {\bf r}}} \mbox{
}\rho ({ {\bf V}}_p \nabla )\mbox{ }\Psi _i
\]

\noindent are the same as the components of Eq. (\ref{eq3}):

\begin{equation}
\label{eq5a}
{\bf
U}^{(0)}= < {\rm {\bf U}}_{tot} > _{A'}.
\end{equation}

The second term is

\[
U_i^{1} = - \frac{G}{c^2}\int {dz} \int {d^2{ {\bf r}}} div({ {\bf
V}_p}\rho \Psi _i ).
\]

For this latter term we have in cylindrical coordinates $\{r,
\varphi, z\}$

\begin{equation}
\label{eq6}
\begin{array}{l}
 U_i^{(\ref{eq1})} = - \frac{G}{c^2}\int\limits_{ - \infty }^\infty {dz}
\int\limits_0^\infty {dr} \int\limits_0^{2\pi } {d\varphi }
\left\{ {\frac{\partial }{\partial r}\left( {r\rho V_r \Psi _i }
\right) + \frac{\partial }{\partial \varphi }\left( {\rho V_r \Psi
_i } \right) +
r\frac{\partial }{\partial z}\left( {\rho W\Psi _i } \right)} \right\} = \\
 = - \frac{G}{c^2}\int\limits_{ - \infty }^\infty {dz} \int\limits_0^{2\pi }
{d\varphi } \int\limits_0^\infty {dr} \left\{ {\frac{\partial
}{\partial r}\left( {r\rho V_r \Psi _i } \right)} \right\} =
\frac{G}{c^2}\int\limits_{ - \infty }^\infty {dz}
\int\limits_0^{2\pi } {d\varphi } \mathop {\lim
}\limits_{r \to 0} \left( {r\rho V_r \Psi _i } \right), \\
 \end{array}
\end{equation}

\noindent where $V_{r}=({\bf V}_p \cdot {\bf e}_r)$,
$V_{\varphi}=({\bf V}_p \cdot {\bf e}_{\varphi })$, ${\bf e}_r=\{
\cos(\varphi ),\sin(\varphi ),0\}$, $ {\bf
e}_{\varphi}=\{-\sin(\varphi ), \cos(\varphi ),0 \}$,  $W$ is the
longitudinal component of ${\bf V}_p $ along  $z$ axis, the latter
is directed along the line of sight to the source.  For  $r
\to 0$ we have $r{\rm {\bf \Psi }} \to - 4{\rm {\bf e}}_r $
($z>0$), and  $r{\rm {\bf \Psi }} \to 0$ ($z<0$). Then simple
calculations yield

\begin{equation}
\label{eq7} {\rm {\bf U}}^{(1)} = -\frac{4\pi
G}{c^2}\int\limits_0^\infty {dz} \mbox{ }\rho (0,z)\mbox{ }{\rm
{\bf V}}_ \bot \quad ,
\end{equation}

\noindent where ${\bf V}_{ \bot }$ represents transverse
components of ${\bf V}_p$. Thus ${\bf U}^{(1)}$ is the difference
of GIM effects for continuous and discrete distributions defined
by mass density on the line of sight. Note that ${\bf U}^{(1)}$
does not go to zero in case of microlens masses fragmentation, if
we make these masses smaller with leaving $\rho $ unchanged.

\section{Average image motion in case of our Galaxy}

We use a four-component model of mass distribution in the Galaxy
(buldge+ disk + halo + dark corona) according to
\cite{MinShal}. We confine ourselves to this model in view of its
simplicity, though more recent and more adequate models of
Galactic density are available at present. However, for
order-of-magnitude estimates the model of \cite{MinShal} is
enough.
 In this model the spherical components (bulge, halo and corona)
  have isothermal sphere mass
density

\[
\rho _S (R) = \frac{3M_S R_S^2 }{4\pi \left( {R_S^2 + R^2}
\right)^{\frac{5}{2}}},
\]

$R$ being distance from the Galactic center.

The parameters of spherical components are taken as follows:
 $M_{B}=1.5\cdot 10^{11}$ (in Solar masses), $R_{B}=5 kpc$;
 $M_{H}=5\cdot 10^{10}$, $R_{H}=25 kpc$;  $M_{C}=8\cdot
10^{11 }$, $R_{C}=50 kpc$. We changed these parameters as compared
to \cite{MinShal} to have better correspondence to the observed
rotation curves.

In case of the Galactic disk (in the cylindrical coordinates
$r,\varphi,z$ with the origin at the Galactic center)

\[
\rho _D (r) = \frac{M_D r_D }{4\pi H\left( {r_D^2 + r^2}
\right)^{\frac{3}{2}}},\mbox{ }\vert z\vert < H;
\quad
\rho _D (r) = 0,\mbox{ }\vert z\vert > H;
\]

\noindent where $H=0.6 kpc$, $R_{D}=15 kpc$ and $M_{D}=8\cdot
10^{10}$.

We use this model along with corresponding rotation curves, which
are recovered by means of potentials of the above mass
distributions \cite{MinShal}. The dependence of the rotation
velocity around the Galactic center in the Galaxy plane is

\[
V^2 = G\left[ {\frac{M_B r^2}{(R_B^2 + r^2)^{\frac{3}{2}}} + \frac{M_C
r^2}{(R_C^2 + r^2)^{\frac{3}{2}}} + \frac{M_H r^2}{(R_H^2 +
r^2)^{\frac{3}{2}}} + \frac{M_D }{2H}\Omega (r)} \right],
\]

\[
\Omega (r) = \frac{2r_D }{\sqrt {r^2 + r_D^2 } } - \frac{r_D + H}{\sqrt {r^2
+ (r_D + H)^2} } - \frac{r_D - H}{\sqrt {r^2 + (r_D - H)^2} }\mbox{ }.
\]

For an observer at the Galactic center we have ${\bf U}_{tot}^\ast
=0$ (see Eq.(\ref{eq4a})); this is evident in case of a stationary
mass density and the observer at rest. We assume  that all the
components except continuous corona consist of stars (though if
the corona also consists of stars, this does not change the result
significantly). Then in view of Eqs.(\ref{eq5}) , (\ref{eq5a}) we
have the average GIM $< { {\bf U}}_{tot}> _{A'}=- {\bf U}^{(1)}$.
This yields for source in the Galactic plane  $3 \cdot 10^{ - 8}$
arcseconds per year. This also can be checked directly by means of
the formula (\ref{eq3}).
When inclination increases, the effect decays strongly due to
decreasing of star number on the line of sight.

\begin{figure}
\resizebox{\hsize}{!} {\includegraphics[]{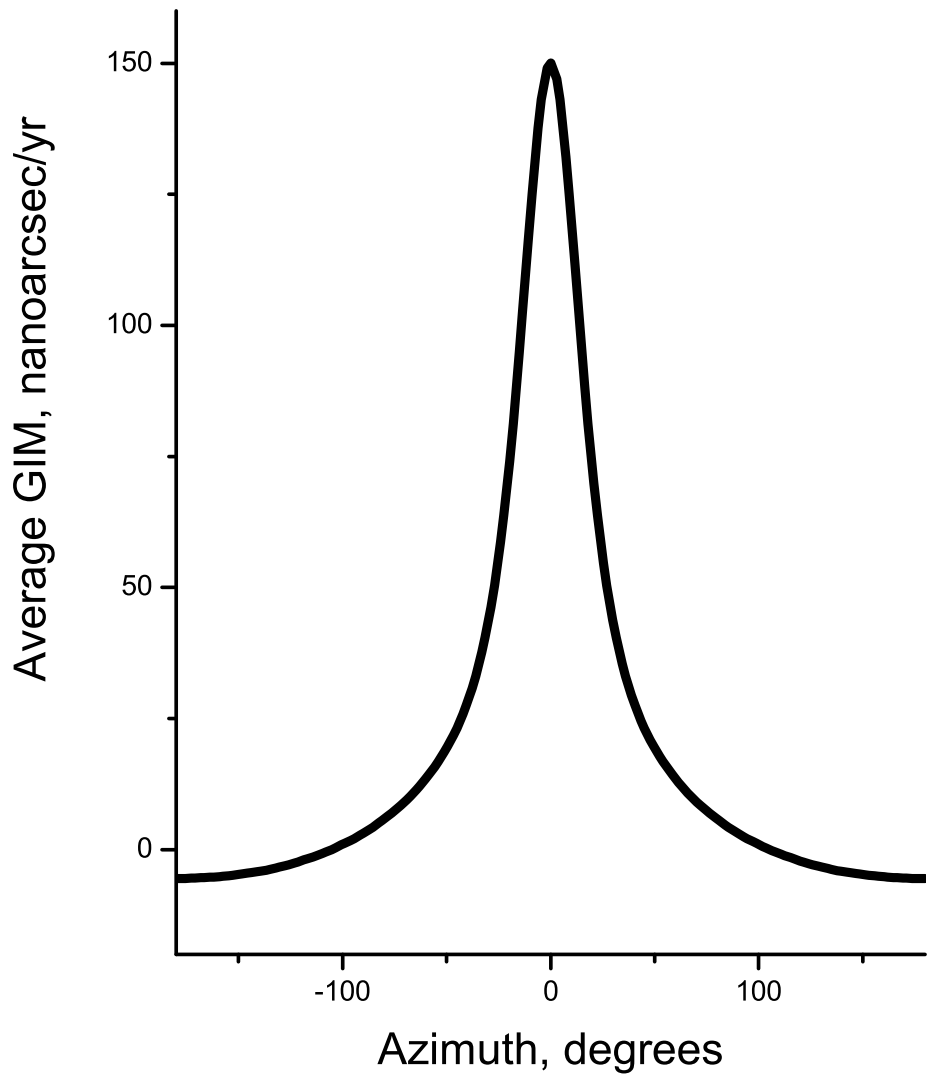}}
\caption{Observer in the Solar system, sources near the Galactic
plane: average GIM against Galactic azimuthal angle.}
\label{label2}
\end{figure}

If the observer is located in the Solar system, we must take into
account its own velocity and the velocity of stellar rotation
around the Galactic center. Note that it is more convenient to
calculate first ${\bf U}_{tot}^\ast $ and ${\bf U}^{(1)}$, then
the value of average GIM $< { {\bf U}}_{tot}> _{A'}$ is defined
from Eq.(\ref{eq5}). Dependence  of GIM upon the Galactic azimuth
of sources in the Galactic plane for the observer in the Solar
system is shown on Fig.\ref{label2}.

\section{Discussion}

We have shown that GIM differs for a continuous and discrete mass
density distribution. In the latter case GIM performs random walks
\cite{ZZ},\cite{Zh},\cite{Sazhin} as distinct from a regular
motion in case of continuous matter. In this paper we draw
attention to the fact that the difference between these two cases
remains also after averaging of the stochastic GIM (for the same
mass density). For example, the observer at the Galactic center
would see an additional fictitious motion of quasars around this
center. However, if most of the Galactic matter would be continuous,
the GIM effect will be absent.

For an observer in the Solar system we have a nonzero average GIM
value leading to apparent rotation of extragalactic reference
frame in the direction of the Galaxy rotation. We stress that the
averaging procedure involves only those realizations of Galactic
stars, which have concern with weak microlensing. For sources in
the Galactic plane maximal effect amounts about $1.5  \cdot10^{ -
7}$ arcseconds per year. This is extremely small. However, in
principle, the effect can be observed, because it falls off with
inclination of the line of sight to Galactic plane. Formal
algorithm for such measurement must involve observations of
sources at different Galactic latitudes. In this procedure the
strongly microlensed images must be eliminated.

As concerned practical measurement of the average GIM effect, we
note that it is far beyond modern possibilities. Moreover,
observation of the average effect requires too many "good"
extragalactic sources with approximately the same Galactic
latitude. This is necessary in order to separate GIM from random
image motions that may be much more essential.

This work is supported by Science and Technology Center in Ukraine
(Project NN43).

\end{document}